\documentstyle[11pt,newpasp,twoside,epsf]{article}
\def\edcomment#1{\iffalse\marginpar{\raggedright\sl#1\/}\else\relax\fi}
\reversemarginpar

\begin{document}
\title{Ultra High Energy Cosmic Ray Accelerators}
 \author{Angela V. Olinto}
\affil{Department of Astronomy \& Astrophysics \& Enrico Fermi
Institute, \\
The University of Chicago, Chicago, IL 60637}

\begin{abstract}
The surprising lack of a high energy cutoff in the cosmic ray spectrum  
at the highest energies together with an apparently isotropic distribution
of arrival directions have strongly challenged most models proposed  for
the acceleration of ultra high energy cosmic rays. Young neutron star
winds may be able to explain the mystery. We discuss this recent proposal
after summarizing the observational challenge and plausible acceleration
sites. Young neutrons star winds  differ from alternative models in the
predictions for  composition, spectrum, and angular distribution which 
will be tested in future experiments.
\end{abstract}

\section{Introduction}

The  detection of cosmic rays with energies above $10^{20}$ eV  has
triggered considerable interest on  the  origin and nature of these
particles. Many hundreds of events with energies above $10^{19}$ eV and
over a dozen events above $10^{20}$ eV  have now been observed by a
number of experiments such as AGASA (Hayashida et al. 1994, Takeda et al.
1998, Takeda et al. 1999), Fly's Eye (Bird et al 1993, 1994, 1995), and 
Haverah Park (Lawrence, Reid,  
\& Watson 1991). Most unexpected is the significant flux of events
observed above $5 \times 10^{19}$ eV (Takeda et al. 1998)  with no sign
of the Greisen-Zatsepin-Kuzmin (GZK) cutoff (Greisen 1966, Zatsepin \&
Kuzmin 1966).  A cutoff should be present if the  ultra-high energy
particles are protons, nuclei, or photons from  extragalactic sources. 
Cosmic ray protons of energies above $5 \times 10^{19}$ eV lose energy to
photopion production off the cosmic microwave background and cannot
originate further than about $50\,$Mpc away from Earth. Nuclei  are
photodisintegrated on shorter distances due to the infrared background
(Puget, Stecker, \& Bredekamp 1976) while the radio background constrains
photons to originate from even closer systems.

In addition to the presence of events past the GZK cutoff, there has
been no clear counterparts identified in the  arrival direction of the
highest energy events. If these
events are protons, they may point back to their sources within a few
degrees, since at these high energies the Galactic and extra-galactic
magnetic fields should  not affect their orbits significantly. The
gyroradius of a $10^{20}$ eV proton is 100 kpc in a
$\mu$Gauss field which is typical for the Galactic disk, therefore,
protons propagate mainly in straight lines as they traverse the Galaxy.
At present, no correlations between arrival directions and plausible
optical counterparts  such as sources in the Galactic plane, the Local
Group, or the Local Supercluster have been found.  The ultra high energy
cosmic ray (UHECR)  data is consistent with an isotropic distributions of
sources in contrast with the anisotropic distribution of light within 50
Mpc from Earth.

\section{The UHECR Puzzle}

In attempting to explain the origin of UHECRs, models confront a number
of challenges. The extreme energy is the greatest challenge that models
for astrophysical acceleration face, and to  complete the puzzle,
models have to match the spectral shape, the primary composition, and
the arrival direction distribution of the observed events.

If UHECRs are extragalactic, the observed  highest energy event at $3
\times 10^{20}$ eV (Bird 1994) argues for accelerators that reach as high
as a ZeV  (ZeV=10$^{21}$ eV).  The energetic requirements at the source
increase with the distance traveled by the UHE primaries from source to
Earth. Depending on the strength and structure of the magnetic field
along the primary's path, the distance traveled may be significantly
larger than the distance to the source.  If  $3 \times 10^{20}$ eV is
taken as a typical energy for protons travelling on straight lines, 
accelerators located further than 30 Mpc need to reach above 1 ZeV while
those located further than 60 Mpc require over 10 ZeV (Cronin
1992). As magnetic fields above $\sim 10^{-8}$ G may thread intragalactic
space (Riu, Kang \& Biermann 1998;  Blasi, Burles \& Olinto 1999),
protons travel in curved paths and sources need  to be either more
energetic or located closer to Earth. 

The extreme energy requirements have encouraged 
alternative explanations for UHECRs. For early universe physics, a ZeV
is not particularly high in energy. For instance, relics from the Grand
Unified scale at $\sim 10^{24}$ eV may be causing these ultra high energy
events (Hill 1983; 
Schramm \&  Hill 1983). The challenge for models that make use of
early universe  relics is generally the flux, the same challenge that
observers face. 
 At 10$^{20}$ eV, the observed flux of UHECRs is
about $\sim$ 1 event/km$^2$/century which has strongly limited our ability to
gather more than a dozen such events after decades of observations.
Although challenging to observers and topological defect models, the flux
is not  particularly constraining in terms of general energetic
requirements on astrophysical sources. In fact, this flux equals the flux
of {\it one} gamma-ray burst that may have taken place in a 50 Mpc radius
volume around us (Waxman 1995; Vietri 1995).  

The energy spectrum of cosmic rays has a
steep  energy dependence  $N(E)  \propto E^{-\gamma}$, with
$\gamma\approx 2.7$  between
$\sim 10^8\,$eV and $\la  10^{15}\,$eV and
$\gamma\approx 3.1$ for $10^{15}$ eV $\la  E \la 10^{19}$ eV.  
Cosmic rays of energy up to  $\sim 10^{15}$ eV are widely accepted to
originate in shocks associated with galactic supernova remnants, but this
mechanism has difficulties producing particles of higher energy.  The
events with energy above
$10^{19.5}\,$eV, however,  show a much flatter spectrum with
$1 \la  \gamma \la 2$. The drastic
change in slope suggests the emergence of a {\it new component} of cosmic
rays  at  ultra-high energies. This new component of cosmic rays is 
generally thought to be   extragalactic, but they may also originate 
in an  extended halo or in the Galaxy
(Olinto, Blasi, \& Epstein 1999) depending on their composition. 

The observed spectrum represents a convolution of the source primary
spectrum with the effect of energy losses during the propagation between
the source and the Earth.  If the new component is extragalactic, loss
processes modify the spectral shape significantly.   For primary protons
the main loss processes are pair production (Blumenthal 1970) and
photopion production off the cosmic microwave background radiation
(Greisen 1966; Zatzepin \& Kuzmin 1966). For straight line
propagation, loss processes limit sources of 10$^{20}$ eV to be within
about 50 Mpc from us. For heavier nuclei, the infrared background
induces  losses (Stecker 1999) and the maximum distance for 10$^{20}$ eV
nuclei to originate from is  $\sim $ 10 Mpc. In the case of photon
primaries,  the radio background represents the main source of losses and
the distance is limited to $\la 10$ Mpc.  Depending on the
poorly known strength and structure of the extragalactic magnetic fields,
the GZK cutoff moves closer in distance for charged primaries. In
addition,  plausible models have to accommodate the spectrum at the
highest energies ($\ga 10^{20}$ eV) without overproducing cosmic rays at 
$\ga 10^{18}$ eV (Berezinsky, Grigorieva, \& Dogiel 1990; Blasi \& Olinto
1998; Sigl, Lemoine, \& Biermann 1998).  

Charged particles of energies up to $10^{20}\,$eV can be deflected
significantly in cosmic magnetic fields. The Larmor radius of a particle
with energy $E$ and charge $Ze$ in a  magnetic field  
$B$,   $r_L= 0.1 {\rm Mpc } (E/10^{20}{\rm eV})/Z (B/\mu G)$.  If the
UHECR primaries are protons, only large scale intergalactic magnetic
fields affect their propagation significantly, while for heavier nuclei
the Galactic magnetic field should also be taken into account. While the
Galactic magnetic field is   reasonably well studied, extragalactic
fields are still very poorly understood.

Together with a composition determination, the distribution of arrival
directions  can hold the key to the UHECR
puzzle. Within a 50 Mpc radius volume around us, the most 
luminous structures are the  Galactic plane, the Local Group and the
general galaxy distribution with a relative overdensity around the region
of the Local Supercluster. If the UHECR source is dark matter, than the
Galactic Halo is the relevant structure which  is expected to be a
spheroidal overdensity  centered around  the Galaxy. On larger
scales the dark matter distribution should correlate with the luminous
matter distribution.  For the few highest energy events, there is
presently no strong evidence for correlations between the events' arrival
direction and any of these known structures (Stanev 1999): the
distribution is isotropic to first approximation.  Since the number of
observed  events above 10$^{20}$ eV is low, it is still early to tell.
For slightly lower energies, some correlations have been detected.  
Recently, the AGASA group has announced that the distribution of
10$^{18}$ eV events shows a significant correlation with the Galactic
Center and the nearby Galactic spiral arms (Yoshida et al 1999). If these
correlations are confirmed, it would be strong evidence for a Galactic
origin for cosmic rays around 10$^{18}$ eV.

\section{UHECR Accelerators}

As mentioned above, there are great difficulties with finding plausible
accelerators for such extremely energetic particles. Even the most
powerful astrophysical objects such as radio galaxies and active galactic
nuclei can barely accelerate charged particles to energies as high as 
$10^{20}$ eV.

\begin{figure}
\plotone{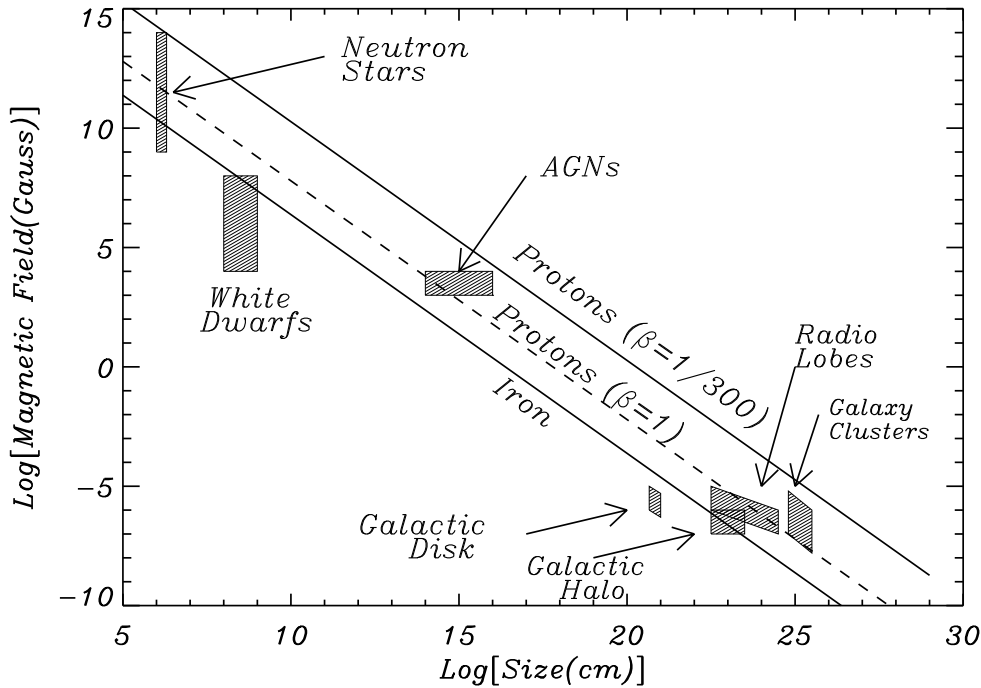}
\caption{B vs. L for E$_{max} = \beta \, Ze \, B \, L$}
\end{figure}

Acceleration of UHECRs in astrophysical plasmas occurs when large-scale
macroscopic motion, such as shocks and turbulent flows, is transferred to
individual particles. The maximum energy, $E_{\rm max}$, is usually
estimated by requiring that the gyro-radius of the particle be contained
in the acceleration region.  Therefore,  $E_{\rm max}$ is usually
associated with the strength, $B$, and coherence length, $L$, of the
magnetic field embedded in the plasma, such that
$E_{\rm max} \sim \beta \, Ze \, B \, L$, where usually $\beta \sim v/c$
and $Ze$ is the charge of the particle. As can be seen in Figure 1, for
$E_{\rm max} \sim10^{20}$ eV and $Z \beta \sim 1$, the only known
astrophysical sources with reasonable  $B L $ products are neutron stars
($B \sim 10^{13}$ Gauss and $L \sim 10$ km),  active galactic nuclei
(AGNs) ($B \sim 10^{4}$ Gauss and $L \sim 10$ AU), radio lobes of active
galaxies ($B \sim 10^{-5}$ Gauss and $L
\sim 10$ kpc), and clusters of galaxies ($B \sim 10^{-6}$ Gauss and $L
\sim 100$ kpc) (Hillas 1984).

 In general, when these sites are considered more carefully, one finds
great difficulties due to either energy losses in the accelerating
regions or the great distances of known sources from our Galaxy. In most
of these objects one invokes shock acceleration as the primary
acceleration mechanism. As discussed by Achterberg in this meeting, shock
acceleration is not effective in reaching ZeV energies for most proposed
accelerators with the possible exception of galaxy cluster shocks. The
problem with clusters of galaxies as sources of UHECRs on Earth is that
our location in the universe is not close enough to a cluster shock to
avoid the GZK cutoff. Furthermore, UHECRs generated in typical clusters
do not escape from them during the age of the universe (Blasi \& Olinto
1998). 

Moving left on Figure 1, radio lobes in FRII objects can  reach the
required energy if $\beta \sim 1$ (Rachen \& Biermann 1993). Again,
location is a challenge since these are rare objects and far apart. A
recent proposal, places the source of all UHECRs observed in a single
object, M87, by invoking a Galactic wind that can make different observed
arrival directions trace back to  M87 (Ahn et al. 1999). The
existence of a Galactic wind  with the required characteristics to allow
for this possibility is yet to be determined. 

Active galactic nuclei (AGNs) are powerful
engines as matter accretes onto very massive black holes. The
problem with AGNs as UHECR sources is two-fold: one is the distance to
more active objects and the other is common among
highly energetic environments - losses due to intense radiation field
downgrades particle energies well below the maximum achievable energy.
These limitations have led to the proposal that remnant quasars, large
black holes in centers of inactive galaxies,  are UHECR accelerators
(Boldt \& Ghosh 1999).  In this case one would have no obvious
counterpart since any galaxy would be as likely to host such
accelerators and losses are not as significant. The spectrum will be
dominated by the local distribution of galaxies with more distance
galaxies inducing a GZK cutoff
 (see, e.g., Medina-Tanco 1999). The
detailed acceleration mechanism for this proposal is yet to be determined.

Before discussing the last possibility  in Figure 1, neutron stars, it
is worth mentioning that the lack of a clear astrophysical solution for
the UHECR puzzle has produced a number of models based on physics beyond
the standard model  such as monopoles, cosmic strings, 
cosmic necklaces, vortons, and superheavy long-lived decaying relic
particles, to name a few. Due to the lack of space,  we refer the
interested reader to a few recent reviews (Berezinsky 1998,
Bhattacharjee \& Sigl 1999).

\subsection{Young Neutron Star Winds}

As shown in Figure 1, neutron stars may be effective in
accelerating UHECRs (see, e.g., Berezinsky et al. 1990). Acceleration
processes inside the neutron star light cylinder are bound to fail
much like the AGN case:  ambient magnetic and radiation fields induce
significant losses  (Venkatesan, Miller, \& Olinto 1997).
However, the plasma that expands beyond the light cylinder is free from
the main loss processes and may be accelerated to ultra high energies.
 
One possible solution to the UHECR puzzle is our recent proposal that the
early evolution of neutron stars may be responsible for the
unexplained flux of cosmic rays beyond the GZK cutoff  (Olinto, Epstein,
\& Blasi 1999; Blasi, Epstein, \& Olinto 1999). In this case,
UHECRs originate in our Galaxy and are due to iron nuclei accelerated 
from  the surface of strongly magnetic young neutron stars.   

Newly formed, rapidly rotating neutron stars may accelerate iron nuclei 
to ultra-high energies  through relativistic MHD winds beyond  their light
cylinders (see, e.g., Michel 1991).  The nature  of the relativistic
wind is not yet clear, but observations of the Crab Nebula indicate that
most of the rotational energy emitted by  the pulsar is converted into
the flow kinetic energy of the particles in the wind. If
most of the magnetic energy in the wind zone is converted into  particle
kinetic energy  and  the rest mass density of the wind is not
dominated by electron-positron pairs,  
 particles  in the wind can reach a maximum energy of
$E_{max} \simeq 8 \times 10^{20}  \, Z_{26} B_{13} \Omega_{3k}^2 \, {\rm
eV}, $ for iron nuclei  ($Z_{26} \equiv Z/26 = 1)$ and neutron star
surface fields $B = 10^{13} B_{13} $ Gauss and initial rotation frequency
$\Omega = 3000 \Omega_{3k}$ s$^{-1}$. In the rest frame of the wind, the
plasma is relatively cold while in the star's  rest frame the plasma
moves with Lorentz factors $\gamma \sim 10^9 - 10^{10}$.

The  typical  energy of the accelerated cosmic rays can
be estimated by considering the  magnetic energy per ion at the
light cylinder $ E_{cr} \simeq B_{lc}^2/ 8 \pi n_{GJ}$ where the
Goldreich-Julian (1969) density is
$ n_{GJ} = 1.7 \times 10^{11} \, {B_{13}  \Omega_{3k}^4 / Z}\, {\rm
cm}^{-3}$. We find $ E_{cr} \simeq  4 \times 10^{20}\,  Z_{26} B_{13}
\Omega_{3k}^2 \, {\rm eV}, $ similar to $E_{max}$  above. 
Therefore, neutron stars whose initial spin periods are shorter than
$\sim 4 (B_S/10^{13}{\rm G})$ ms can accelerate iron nuclei to
greater than $10^{20} eV$.  

\begin{figure}
\plotone{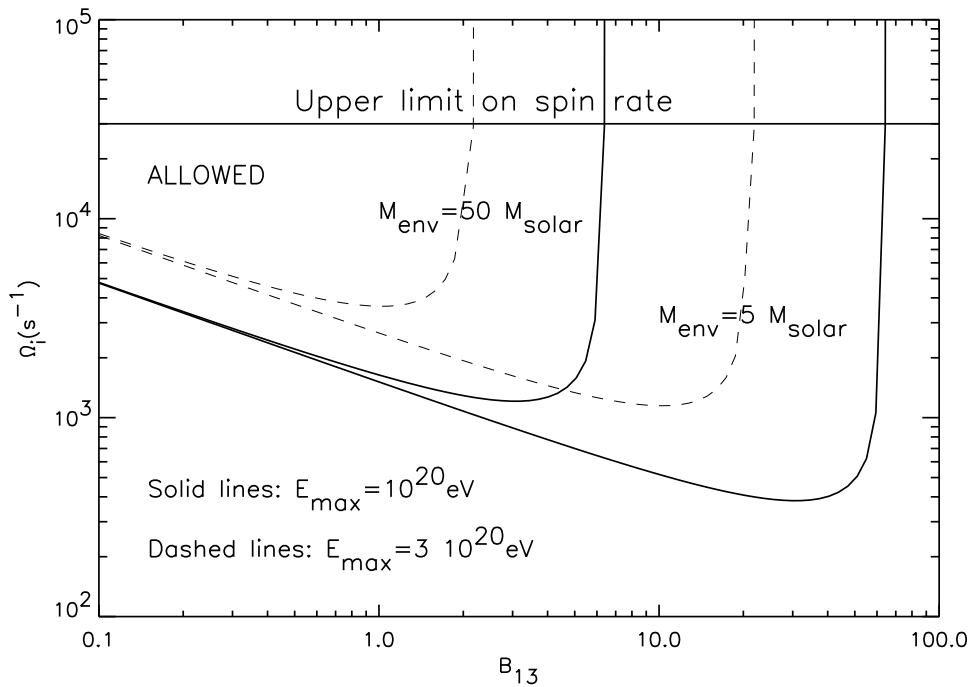}
\caption{Solid
lines  for  $E_{cr}=10^{20}$ eV and dashed lines
for  $E_{cr}= 3\times 10^{20}$ eV. The curves are plotted for two
values of the envelope mass, $M_{env}=50~M_{solar}$ and
$M_{env}=5~M_{solar}$, as indicated. The horizontal line at spin
period $\sim 0.3$ ms indicates the minimum period allowed for neutron
stars.}
\end{figure}
 
About a year  after the supernova explosion,
the iron nuclei can escape through the remnant of the supernova that
produced the neutron star without suffering significant spallation
reactions.  The supernova event  ejects the envelope of the original star,
making it possible for cosmic rays to escape. However, as the envelope
expands,  the young neutron star spins down and may become unable to emit
particles of the necessary energy. A requirement for relativistic winds
to  supply  UHECRs is that the column density of the envelope becomes
transparent to UHECR iron 
before the spinning rate of the neutron star decreases significantly.
The allowed parameter space for this model
is shown in Figure 2. 

The spectrum of accelerated UHECRs is determined by the evolution of
the  rotational frequency: As the  star spins down, the energy of
the cosmic ray particles ejected with the wind decreases.
 The predicted spectrum 
 is very flat, $\gamma =1$,  in   agreement with the UHECR
data. Furthermore, for the parameters within the allowed region, the
acceleration and survival of UHECR iron nuclei is not significantly
affected by the ambient photon radiation.

Depending on the structure of the  galactic
magnetic field, the trajectories of the iron nuclei from galactic
sources can be consistent with the observed arrival directions of the
highest energy events (Zirakashvili et al. 1998).
The gyroradius of these UHECRs  in the Galactic
field of strength $B_{gal} = 3 \mu$ Gauss is
$ r_B = 1.4 {\rm \ kpc} E_{20} /  Z_{26} $ which is considerably less than
the typical distance to a young neutron star ($\sim $ 8 kpc). Furthermore,
the cosmic ray component at
$10^{18}$ eV is nearly isotropic. If these cosmic rays are protons of
Galactic origin, the isotropic distribution observed
at these energies may be indicative of the diffusive effect of the
Galactic and halo magnetic fields. Iron at
$10^{20}$ eV probes  similar trajectories to protons at a few times
$10^{18}$ eV.

\section{Conclusion}

Future experiments such as the Auger Project and the OWL-Airwatch
satellite will be able to discriminate between different models (Cronin
1999; Watson 1999). An excellent discriminator would be an unambiguous
composition determination  of the primaries. In general, Galactic disk
models  invoke iron nuclei to be consistent with the isotropic
distribution of events, and extragalactic astrophysical
models tend to favor proton primaries, while photon primaries are more
common for early universe relics. The observational tools in place for
composition discrimination are the muon content of shower in the ground
arrays  (more muons for nuclei vs. nucleons) and the depth of shower
maximum in fluorescence detectors (the heavier the primary the  deeper in
the atmosphere their shower maximum).   In addition, the  correlation of
arrival directions  for events with energies above
$10^{20}$ eV  with some known structure would be key in differentiating
between different models. For example, we should see a correlation with
the Galactic center and  disk for the case of young neutron star winds
(see, Stanev 1999; Yoshida et al.  1999), and the large scale galaxy
distribution for the case of quasar remnants. Both aspects will be
testable with future experiments.

\acknowledgments  
We thank the organizers and the  support of NSF
through grant AST 94-20759  and DOE through grant DE-FG02-90ER40606 at
the  University of Chicago.

\end{document}